\documentclass[fleqn,10pt]{olplainarticle}

\usepackage{tabularx}

\title{A Coupled Two-Tier Mathematical Transmission Model to Explore Virulence Evolution in Vector-Borne Diseases}

\author[1]{Daniel A.M. Villela}
\affil[1]{Program of Scientific Computing, Oswaldo Cruz Foundation, Rio de Janeiro, Brazil\\
Email: daniel . villela@fiocruz.br.}


\begin{abstract}

The emergence or adaptation of pathogens may lead to epidemics, highlighting the need for a thorough understanding of pathogen evolution.
The tradeoff hypothesis suggests that virulence evolves to reach an optimal transmission intensity relative to the mortality caused by the disease.
This study introduces a mathematical model that incorporates key factors such as recovery times and mortality rates, focusing on the diminishing effects of parasite growth on transmission, with a focus on vector-borne diseases.
The analysis reveals conditions under which heightened virulence occurs in hosts, indicating that these factors can support vector-host transmission of a pathogen, even if the host-only component is insufficient for sustainable transmission.
This insight helps explain the significant presence of pathogens with high fatality rates, such as those in vector-borne diseases.
The findings underscore an elevated risk for future outbreaks involving such diseases.
Enhanced surveillance of mortality rates and techniques to monitor pathogen evolution are vital to effectively control future epidemics. This study provides essential insights for epidemic preparedness and highlights the need for ongoing research into pathogen evolution.

\end{abstract}

\definecolor{color2}{RGB}{255,255,255} 

\begin{document}

\flushbottom
\maketitle
\thispagestyle{empty}


\section*{Introduction}

Preparing for potential pathogen emergence or re-emergence requires a comprehensive understanding of the evolutionary trends of pathogens. 
Despite advancements in various treatment approaches, infectious diseases continue to claim tens of millions of lives over the years.
The Global Burden Study reported that more than 13 million deaths related to infected syndromes occurred in 2019 \citep{ikuta_global_2022_short}. In terms of viruses, more than 200 viruses are known to infect humans \citep{woolhouse_temporal_2008}. In addition, there are also non-viral infectious diseases.  
%
For over a decade, the World Health Organization (WHO) has actively addressed the emergence of new diseases, recognizing them as significant threats requiring thorough preparation and planning \citep{knobler_learning_2004}, which should include vector-borne diseases.  
Understanding virulence patterns through theoretical frameworks can assist in comprehending various scenarios and adjusting preparedness levels accordingly.

In particular, vector-borne diseases are well-characterized by the alternate transmission between vectors and hosts. 
Deaths by vector-borne diseases are critical in the current infectious diseases realm,
contributing to over 17\% of deaths \citep{organization_global_2017}.
\cite{ewald_host-parasite_1983} found that among 45 diseases without vectors, only 5 had a fatality rate above 1\%, whereas for 18 vector-borne diseases, 8 had a fatality rate above 1\%.
These vector-borne diseases include malaria, with approximately 576,000 cases reported in 2019 across several continents
\citep{world_health_organization_world_2023}. 
Noteworthy, about 95\% of 
deaths due to malaria occurred in African countries in 2021 \citep{world_health_organization_world_2023}.
%

The tradeoff hypothesis \citep{bolker_transient_2009} on disease evolution posits a tendency towards lower morbidity and reduced mortality, driven by the notion that more virulent pathogen strains could hinder transmission due to smaller host pools or host absence resulting from fatalities.
By contrast, pathogens with negligible morbidity and mortality are likely to have very low transmission rates, jeopardizing their persistence.
The most likely scenarios for pathogens in the evolutionary landscape are in achieving a delicate balance for their sustained maintenance.
%
In fact, there is a significant literature \citep{dieckmann_adaptive_2002} that discusses that virulence may evolve to maximize the transmissibility compared to the recovery rate and mortality rates in the hosts.  These theoretical frameworks consider the maximization of the basic reproduction number.

Formal studies often assess the optimal mortality level for pathogens to maximize the basic reproduction number, assuming diminishing effects on transmission as virulence increases.
%
%
Vector-borne diseases, which involve at least two levels of transmission (host-vector-host), have received surprisingly little attention in theoretical frameworks of virulence.
This study presents a comprehensive mathematical model extending a general compartmental model to multiple levels. The model incorporates incubation periods, recovery time, and mortality, all dependent on the parasite growth rate, demonstrating diminishing effects on transmission. The model is designed to describe general transmission dynamics in mosquito-borne diseases and can be applied to other vector-borne diseases.
%

The formal model enables the determination of the optimal combination of virulence levels in both vectors and hosts, revealing conditions in the state space where high virulence occurs in hosts. In such cases, a single level for host-only transmission may be insufficient for effective transmission, potentially explaining the higher prevalence of pathogens with elevated fatality rates among vector-borne diseases. 

\section*{Mathematical Model and Analysis}


\subsection*{Derivation vector-host SEIRD model}

The model assumes coupled SEIRD compartmental models for disease transmission dynamics in both hosts and vectors. 
The model includes general parameters for mosquito-borne diseases, such as incubation times, mortality rates, disease-induced mortality, and transmission factors in both humans and vectors.
Transmission parameters for hosts and vectors are denoted by $\beta_h$ and $\beta_v$, respectively. The average incubation period is given by $\theta_h^{-1}$ for hosts and $\theta_v^{-1}$ for vectors. Hosts experience natural mortality $\mu$, while vectors have a natural mortality rate $g$. Disease mortality parameters are denoted by $\omega$ and $\delta$ for hosts and vectors, respectively.
Equations, normalized by total population sizes, describe the dynamics for susceptibles ($S_x$), exposed ($E_x$), infected ($I_x$), recovered ($R_x$), and dead individuals ($D_x$), where $x$ can be either $h$ for hosts or $v$ for vectors.
The transmission intensity terms are functions of virulence in both hosts and vectors. The number of newly infected hosts (vectors) at time interval $dt$ depends on susceptible hosts (vectors), infected vectors (hosts), and transmission intensities $\beta_h$ and $\beta_v$ for hosts and vectors, respectively.  Transmission intensities are given as functions of disease mortality $\beta_h(\delta, \omega)$ and $\beta_v(\delta, \omega)$ in an abuse of notation.

The disease cycle involves alternating phases of transmission involving vector and hosts, i.e, the vector--host transmission. Analyzed separately, each of these phases permits a theoretical construct that establishes the reproduction number for host-only $R_{0,h}$ and vector-only $R_{0,v}$ transmissions.
When taking into account the host--vector--host transmission, the reproduction number is the geometric mean between the reproduction numbers for these two phases:
\begin{equation}
    R_0 = \sqrt{R_{0,h} m R_{0,v}},
\end{equation}
where the ratio $m$ between abundances of host and vector is clearly an amplifying factor.
The complete set of equations of the model is provided in the appendix. The theoretical framework for analyzing the basic reproduction number involves the matrix-generation method.

\subsection*{Conditions for the optimal state}

An evolutionary trajectory is likely to lead to a virulence level that maximizes the reproduction number.
The theoretical methodology for studying optimal levels of virulence involves solving equations for partial derivatives with respect to virulence variables of hosts and vectors, expressed as follows:
\begin{align}
\label{eq:partials}
\quad \partial_{\delta} R_0(\delta, \omega) & = 0 \mbox{ and } 
\quad \partial_{\omega} R_0(\delta, \omega)  = 0,
%
\end{align}
where the notation $\partial_\delta = \frac{\partial R_0(\delta, \omega)}{\partial \delta }$ and $\partial_\omega = \frac{\partial R_0(\delta, \omega)}{\partial \omega }$ is used for conciseness.  

\subsection*{Transmission rate as function of mortality rate}






The first step to define the transmission rate in a single SEIRD model is to describe the replicating dynamics within an infected individual.
The transmission model is such that the number of parasites in a host increases over time with kinetics given by an increasing function $N(t)$.
The function considered in the model is $N(t) = A(1-e^{-r t})$, where $A$ is a ceiling constant and $r$ is the parasite replicating rate.
There are significant properties in this formulation. First, the ceiling factor means that there is a maximum number of parasites in the host.  Second, as density increases, we expect the growth will slow down. 
Typically, the time $t_d$ to reach a deadly amount $A_d$ is such that $ A(1-e^{-r t_d}) = A_d$.
Hence,
$t_d = 1/\omega = -\log(1-A_d/A)/r$.
%
Hence, rate $r$ is proportional to mortality rate $\omega$, and is convenient to write $r = k \omega$, where 
$k=(-1/\log(1-A_d/A))$.

The transmission rate is intuitively how many transmission events occur over time.  The number of events is a random variable $\Omega$, for which the mean value of its distribution will be of interest here.  The expected number of transmissions over a time $\tau$ is $N(\tau)/N_0$, where $N_0$ is the quantity of parasites required for a single transmission. Formally, the transmission rate conditioned on the time $\tau$ is $E[\Omega|\tau] = \frac{N(\tau)}{N_0 \tau}$, where time $\tau$ in the denominator effectively indicates the number of events per time $\tau$.  

The probability of transmission in time $\tau$ depends on the time $T_{min}$, a random variable given by the minimum value between the recovery time $T_r$, host death time $T_d$. Assuming an exponential distribution for these times, the distribution of $T_{min}$ is also exponential:
$$P(T_{min} > t) = P(T_m > t)P(T_r > t) = e^{-\mu t} e^{-\gamma t},$$ where $\mu$ and $\gamma$ are general mortality and recovery rates in the distributions of $T_m$, and $T_r$. Hence, the distribution of $T_{min}$ is exponential with a rate given by the sum of these rates. 

Finally, the transmission rate $\beta_h(\omega)$, as a function of mortality $\omega$, is given by $E[\Omega] = \int_0^{\infty} E[\Omega | \tau] P(T_{min} = t) dt$. 
Using the formulation of the replicating dynamics,  the overall transmission rate $\beta(\omega)$ is
$$
\beta(\omega) = \int_0^{\infty} ([A(1-e^{-k \omega \tau})]/N_0 \tau) (\mu+\gamma) e^{-(\mu+\gamma) \tau} d\tau = 
%
-\Omega_c \log(\mu+\gamma) + \Omega_c \log(\mu+\gamma+ k \omega),$$ where $\Omega_c$ is the constant given by $A(\mu+\gamma)/N_0$.  
This formulation is also applied to vectors for finding $\beta_v(\delta)$ with the appropriate change to vector parameters.




\subsection*{Parameters for scenario comparisons}

There is significant literature on the survival of vectors and hosts. The host parameters for scenario comparison purposes concentrate on human populations to compare different outcomes in human populations.  For hosts, mortality can be sourced from well-studied diseases impacting the human population. Life expectancy varies significantly across different countries, with an overall global life expectancy of close to 70 years in 2021. Consequently, $\mu = (70 \times 365)^{-1}$ (day$^{-1}$).
The survival parameters for vectors can vary with species.  For mosquito species, capture and recapture studies revealed good estimations of the survivorship of mosquitoes in different settings apart from the laboratory conditions.
%
A baseline value for mortality of \textit{Ae. aegypti} is $-\log(0.8)=0.097$ per day, estimated in capture-recapture studies with \textit{Ae. aegypti} mosquitoes \citep{villela_novel_2017}. However, this mortality rate can vary widely in the field depending on environmental conditions, whereas {\em Aedes} mosquitoes can survive a few weeks in laboratory conditions.
A contrasting survivorship is a key difference between most host--vectors.
The rate $g$ was studied as either given by 5 days of survival time or 20 days of survival time.
The recovery rate for hosts is $1/5$ day$^{-1}$ and for vectors is 0.05 day$^{-1}$.
The mean incubation rate is 0.2 for vectors and 0.5 day$^{-1}$ for hosts. 
The constant $\Omega_c$ is 8.3 $\times 10^{-4}$ day$^{-1}$ for vectors and 1.6 $\times 10^{-3}$ day$^{-1}$ for hosts. The abundance is set initially at $m=1$ for equal treatment purposed, and subsequently, experimented with a multiplying factor for vectors.


\subsection*{Comparison of coupled vector--host, and host--only, and vector--only components}

The basic reproduction number for a single tiered SEIRD compartmental model is derived in the appendix.
The basic reproduction number in the two-tiered model is a product between the reproduction numbers due to vectors and hosts:
\begin{equation}
\label{eq:components}
    R_0(\delta, \omega) = \sqrt{m R_{0,v}R_{0,h}},
\end{equation}
where $R_{0,v}$ is the reproduction number factor due to vectors and $R_{0,h}$ is the reproduction number factor due to hosts.
The main criterion for the viability of sustainable disease transmission to be evaluated is the conditions for virulences levels given by $\delta$ and $\omega$ in the vectors and hosts such that:
%
 $   R_0(\delta, \omega) > 1$.

Given the parameters of the vector-host model, the  vector-only component has only the vector--related factors in the transmission. Similarly, the host--only component has the host factor in the reproduction number. 
The transmission terms in these components are according to the transmission formulation: 
\begin{align}
\beta_h(\omega) = \Omega_c( \log(k_h \omega + \mu + \gamma_h) - \log(\mu + \gamma_h)) \mbox{ and }
\end{align}
\begin{align}
\beta_v(\delta) = \Omega_c( \log(k_v \delta + g + \gamma_v) - \log(g + \gamma_v) ).
\end{align}


Given these model counterparts, the definitions for $R_{0, v}$ and $R_{0, h}$ give the reproduction numbers under these components. For the SEIRD component with the vector, the factor in the reproduction number is given by $ R_{0,v} = \frac{\theta_v \beta_{v}(\delta)}{(\theta_v + g)(g+\gamma_v(\delta) + \delta)}$.





%
%
%


\section*{Results}

\subsection*{Virulence that maximizes the reproduction number}

The matrix-generation method for finding $R_0$ requires analyzing the infectious states $E_v, I_v, E_h, I_h$ in the mathematical model. The reproduction number as a function of mortality parameters $\omega$ and $\delta$ is
%
%
\begin{equation}
\label{eq:R0formula}
R_0(\delta, \omega) = \sqrt{m \frac{\beta_v(\delta, \omega) \theta_v(\delta)}{(g + \theta_v(\delta))(g+ \delta + \gamma_v(\delta))} \frac{\beta_h(\delta, \omega) \theta_h(\omega)}{(\mu + \theta_h(\omega))( \mu + \omega + \gamma_h(\omega))}}.
\end{equation}

The reproduction number $R_0$ 
is composed of a product of factors given by vector transmission, host transmission, and the number of vectors per hosts. Equation \ref{eq:R0formula} can be written with factors as formulated in Equation \ref{eq:components}, 
\begin{align*}
R_{0,h} & = \frac{\beta_h(\delta, \omega) \theta_h(\omega)}{(\mu + \theta_h(\omega))( \mu + \omega + \gamma_h(\omega))} = \frac{\Omega_h(\log(k_h \omega + \mu + \gamma_h) - \log(\mu + \gamma_h))}{\mu + \gamma_h + \omega} \\
R_{0,m} & = \frac{\beta_v(\delta, \omega) \theta_v(\delta)}{(g + \theta_v(\delta))(g+ \delta + \gamma_v(\delta))}  = \frac{\Omega_v(\log(k_v \delta + g + \gamma_v) - \log(g + \gamma_v))}{g + \gamma_v + \delta}.
\end{align*}



If $\beta_v(\delta, \omega) = \beta_v(\delta)$ for any $\omega$ and
$\beta_h(\delta, \omega) = \beta_h(\omega)$ for any $\delta$, i.e. transmission in one host does not depend on mortality in the other hosts, then
$\partial_\delta R_0 = 0$, and also implies that
$\partial_\delta R_{0,v} =0$, 
$\partial_\omega R_0 = 0$, and
$\partial_\omega R_{0,h} =0$.
Clearly, the optimal solution for $R_0^2$ will also be solution for $R_0$.
Typically the solutions for $\partial_\delta R_{0,v} =0$ require 
$$\frac{\partial_\delta \beta_v(\delta)}{\beta_v(\delta)} = \frac{1}{g + \delta + \gamma_v}.$$
Therefore, the condition for maximization of $R_0$ is such that the logarithmic derivatives of numerator and denominator are equal. 
For the conditions of constant $\theta_v$, $\theta_h$, $\gamma_v$ and $\gamma_h$ we have 
\begin{equation*}
\frac{1}{(g+\gamma_v+k \delta)(\log(g+\gamma_v+\delta) - \log(g+\gamma_v)} = \frac{1}{g+\gamma_v + \delta}
\end{equation*}

Therefore, optimal mortality rate $\delta_{opt}$ is the solution to the equation:
\begin{equation*}
(g+\gamma_v+ k_v \delta_{opt})(\log(g+\gamma_v+\delta_{opt}) - \log(g+\gamma_v)) = g+\gamma_v + \delta_{opt},
\end{equation*}
which can be solved numerically.
A similar line of reasoning leads to the result for the optimal mortality rate, $\omega_{opt}$, in the host:
%
\begin{equation*}
(\mu+\gamma_h+ k_h \omega_{opt})(\log(\mu+\gamma_h+\omega_{opt}) - \log(\mu+\gamma_h)) = \mu+\gamma_h + \omega_{opt}.
\end{equation*}

\begin{figure}[ht]
\centering
\includegraphics[width=0.8\linewidth]{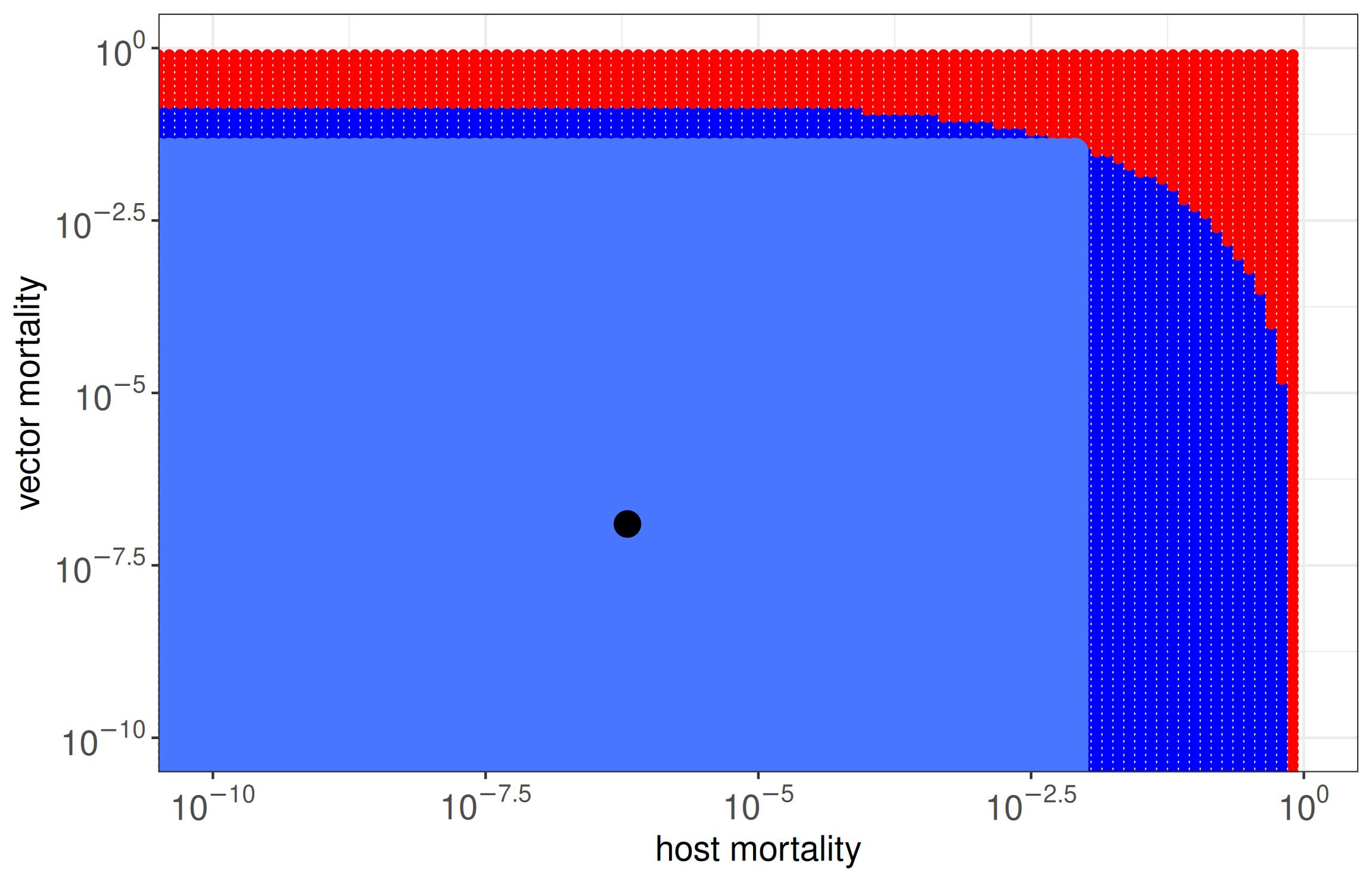}
\caption{The space of state conditions for viable transmission varying by mortality  rates in logarithmic scale, given by vector mortality (y-axis) and host mortality (x-axis). The area in which both $R_{0,v}>1$ and $R_{0,h} > 1$ is given by the light-blue area. The solid light-blue area shows the area given in which $R_{0} > 1$ by simultaneous conditions of the product between $R_{0,m}$ and $R_{0,h}$.  The red area by contrast, shows the mortality conditions in which transmission does not occur. 
The black dot shows the coordinates for highest $R_0$.
}
\label{fig:view}
\end{figure}

\subsection*{A vector-host mode expands the virulence space with viable transmission}

The condition for $R_{0}^2>1$ is given by
$$m \frac{\beta_v(\delta, \omega) \theta_v}{(g + \theta_v(\delta))(g+ \delta + \gamma_v(\delta))} \frac{\beta_h(\delta, \omega) \theta_h}{(\mu + \theta_h(\omega))( \mu + \omega + \gamma_h(\omega))} > 1.$$
There are conditions under which the vector-only component is less than one and the basic reproduction number for the complete vector--host transmission is above 1.




\begin{figure}[ht]
\centering
\includegraphics[width=.65\linewidth]{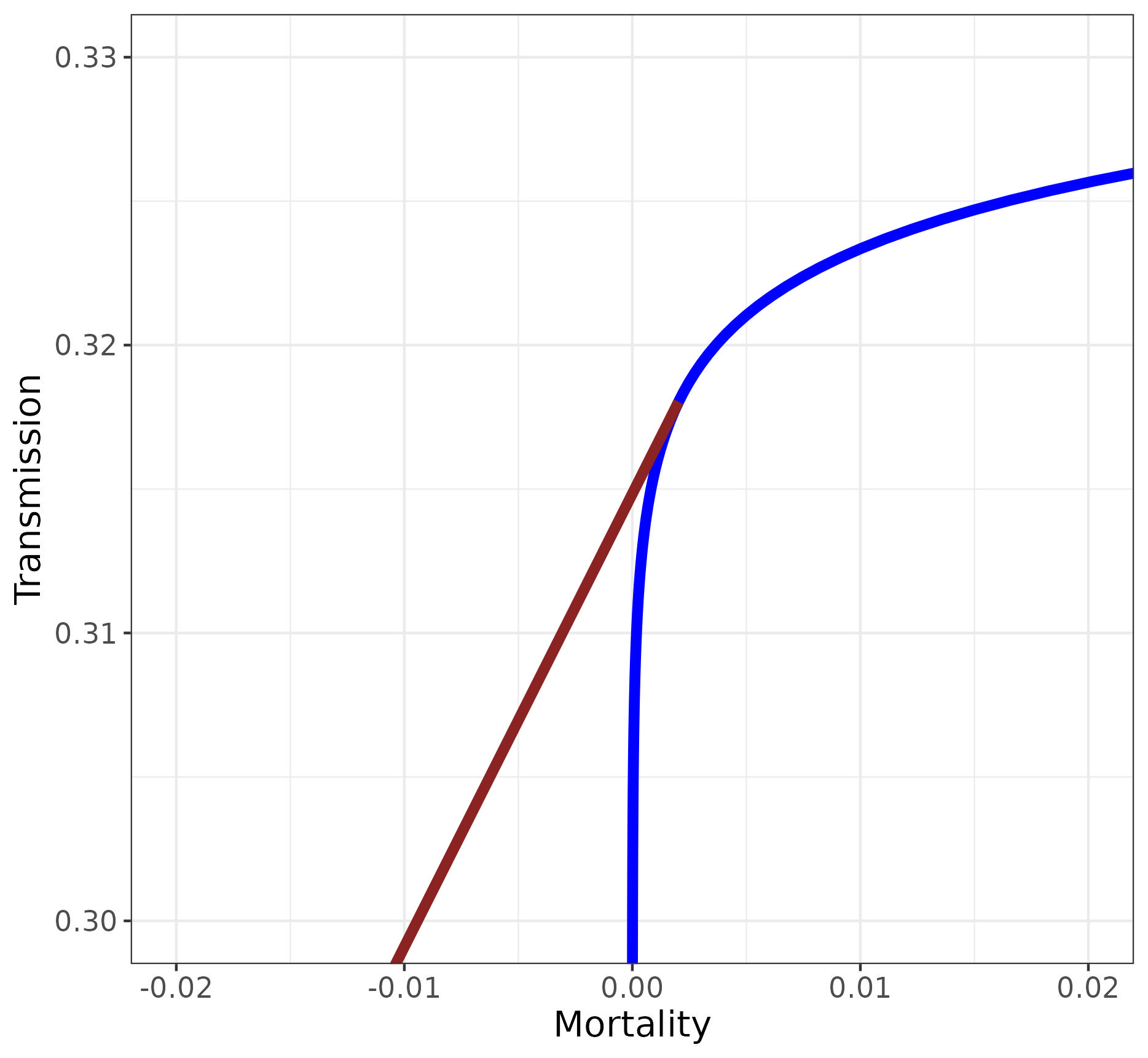}
\caption{The transmission rate (blue color) as a function of mortality for values $\omega > 0$.
The brown color is constructed for illustration purposes with a line from $(-\mu + \gamma_h, 0)$ to the point of $(\omega, \beta_h(\omega))$ for which reproduction number is maximum.
}
\label{fig:maxR0}
\end{figure}

Figure \ref{fig:view} shows how the vector--host transmission expands the area in the viability space of mortality rates. Viability is evaluated as $R_0>1$.
The solid-blue area expands the viability area both for host and vectors, as more conditions of  however the effect is greater for hosts.
The horizontal expansion given by the dark blue area shows the levels of host mortality that can be reached with outbreak conditions ($R_0>1$).  The vertical expansion of the dark blue area shows the levels of vector mortality that can be reached with $R_0>1$.



The theoretical framework presented by \cite{dieckmann_adaptive_2002} illustrates for a generic directed-transmited disease a function of transmission in a cartesian plot, such that the ratio given by $\beta$ and the virulence plus natural mortality and recovery times should the maximized.  In that case, the line from $x= - (\mu + \gamma_h)$ to the points in the plot of $\beta_h$ should be such that the line is tangent to the curve for an optimal point.  In this case, the same reasoning applies, as shown in Figure \ref{fig:maxR0} with the same parameters from Figure \ref{fig:view}. 
 
\begin{figure}[ht]
\centering
\includegraphics[width=0.65\linewidth]{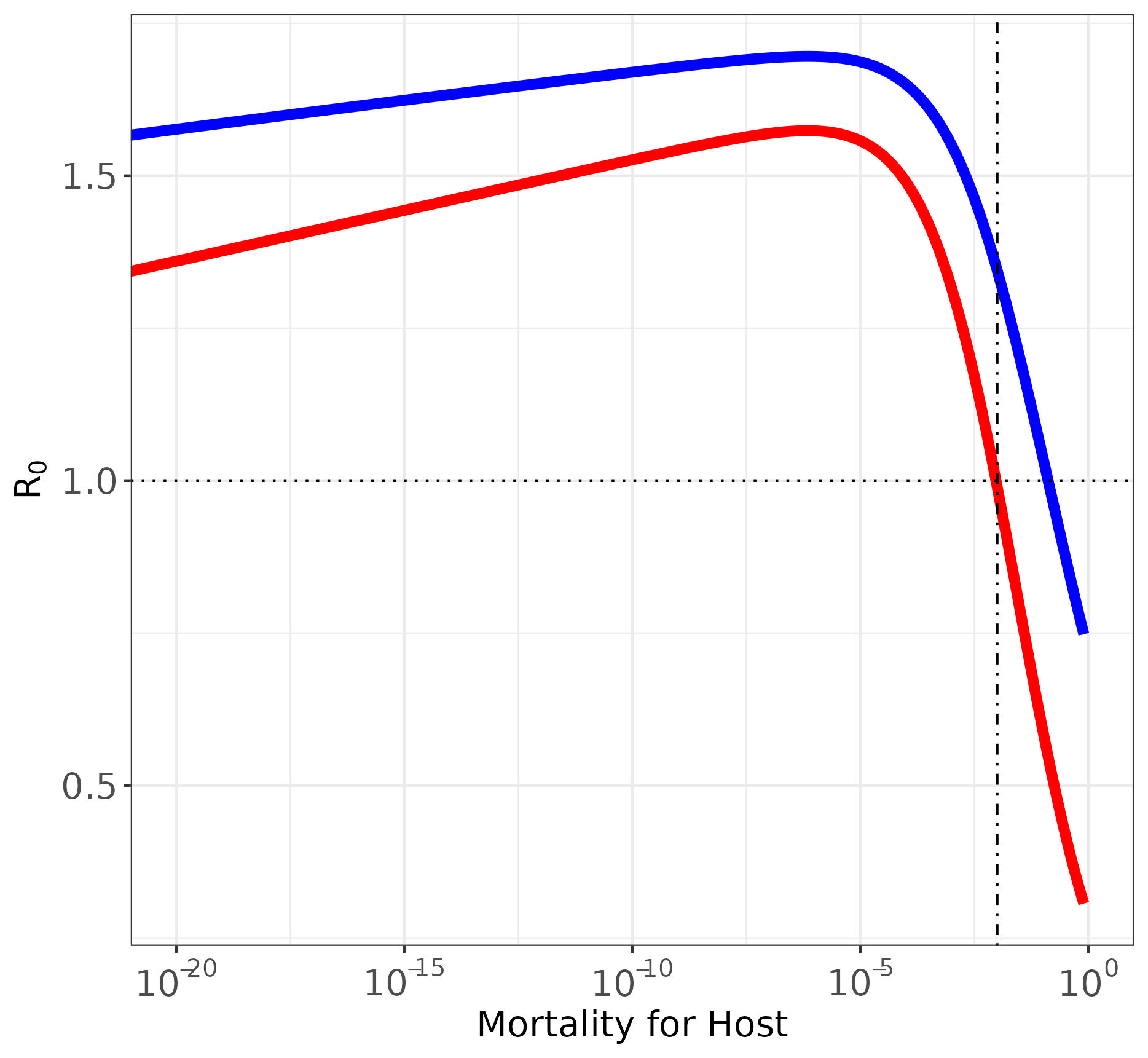}
\caption{The basic reproduction number $R_0$ as a function of mortality (logarithmic scale).  The top curve (blue) shows the reproduction number for complete vector--host transmission, wheres the lower curve (red) shows 
the host-only component of the reproduction number.  The horizontal line is shown as a reference for $R_0=1$ and the vertical line shows the mortality at $10^{-2}$. 
Illustration of optimal state.}
\label{fig:graphR0}
\end{figure}


\begin{figure}[ht]
\centering
\includegraphics[width=0.7\linewidth]{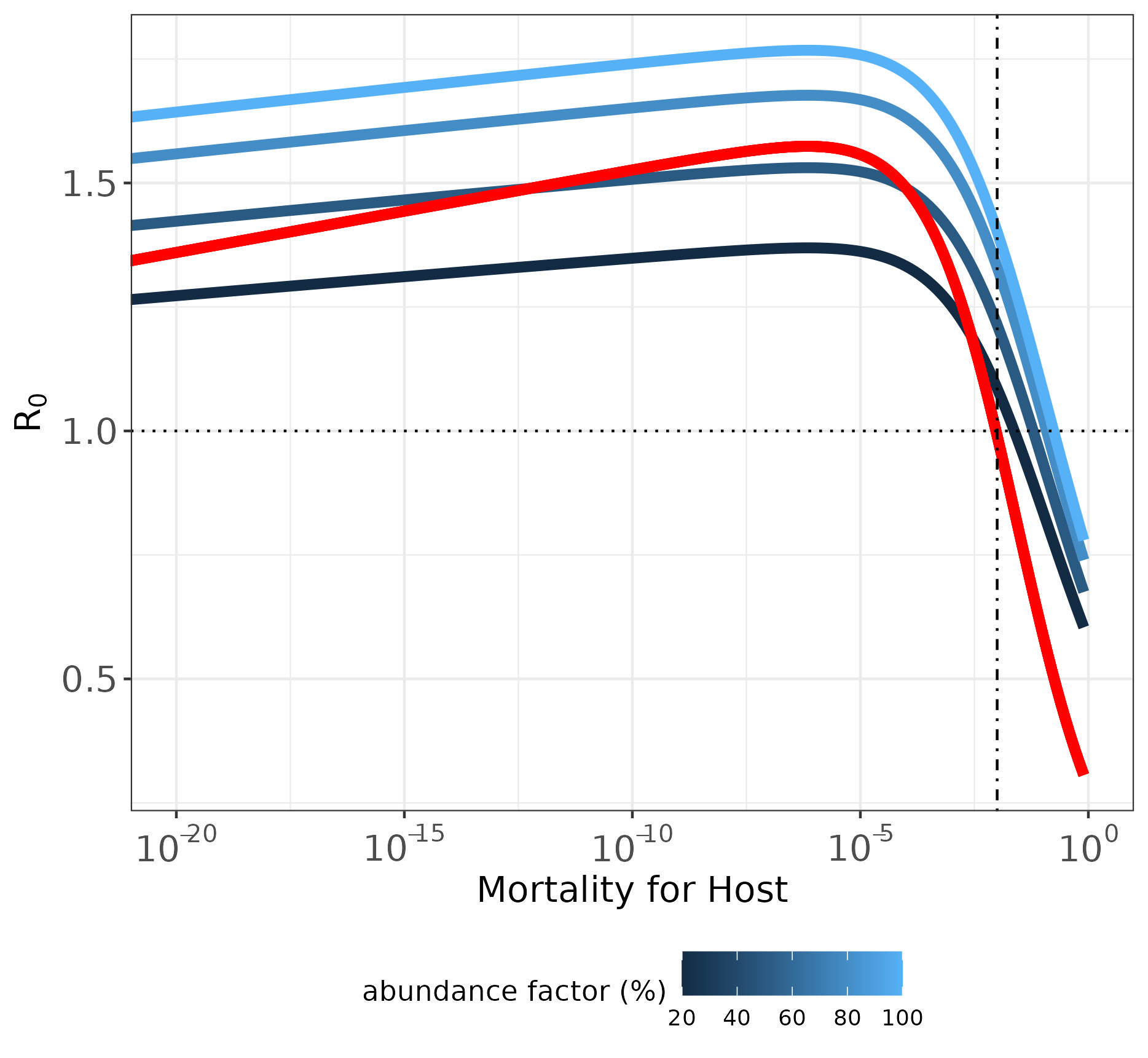}
\caption{The reproduction number as a function of the host mortality for host-vector transmission (dark blue to light blue) and in the host--only transmission (red color).  The horizontal dashed line shows the line where $R_0=1$ for reference purposes.  The vertical dashed line shows the mortality rate in which host-only $R_0=1$.}
\label{fig:graphR0_abund}
\end{figure}



Figure \ref{fig:graphR0} shows how the reproduction number can be higher for a host-vector transmission than the host-only reproduction number.  The sufficient condition for a higher transmission in host-vector transmission is that the vector component is higher than the host component, given that the joint reproduction number is given by the geometric mean.  In this setting $R_{0,h} = 1$ for $\delta = 10^{-2}$, whereas the joint $R_0 > 1.2$ as seen by the dashed vertical line.

The vector can also contribute to transmission by abundance.  In fact, the parameter $m$ might profoundly impact.  Figure \ref{fig:graphR0_abund}
 shows that the abundance might make the vector--only transmission higher or lower than the host--only transmission as the abundance factor varies from 20\% to 100\%.  As the abundance factor increases by a factor of 1.8, the joint reproduction number is clearly above the host--only reproduction number for the range of mortality rates.

Finally, the appendix presents an independent evaluation of the fatality rates of several diseases studied in the 2019 GBD study \citep{institute_of_health_metrics_and_evaluation_global_2020}.
Several vector-borne diseases have significant fatality rates, especially in the group of low--income countries. 
The fatality rate can typically be given as a function of the disease mortality using a SEIRD model, asymptotically reaching  $\frac{w}{w+\gamma}$, similar to the derivation by \cite{carcione_simulation_2020}.
 
\section*{Discussion}



The theoretical framework in this study shows that interactions between vector and host transmission position vector-borne diseases within a wide range of virulence states that enable transmission. 
The vector component may amplify the reproduction number increasing the evolutionary viability for the pathogen and may widen the range of mortality values in which this viability happens. 
This can lead to increased virulence in hosts, even if it does not  correspond necessarily to the highest reproduction numbers. The varying levels of virulence can have serious consequences; for example, yellow fever is known to have a high fatality rate \citep{douam_yellow_2018}. These findings highlight the need for heightened attention to vector-borne diseases as potential emergent threats.

%
%

The evolutionary state of current known pathogens spreads across a broad spectrum of mortality rates, as disease mortality rates for known diseases range from mild to very high.
Therefore, the current severity state may not be the optimal, i.e., the one that maximizes the reproduction number for a given pathogen. 
The reproduction number can be higher than the one, as a theoretical condition for sustainable transmission, provided by the host--only component, given that the vector component introduces a factor that amplifies the overall reproduction number.  In the formulation of the reproduction number the interaction of these components is given by the geometric mean.
In particular, a reproduction number above 1 for a given state whereas the host--only component is below 1 signals a viability enabled by the vector--host transmission.
%
%

Geographical and climate conditions may restrict vector distribution, meaning they might not be present globally. 
\cite{de_angeli_dutra_evolutionary} also discuss the possibility of higher virulence in vector-host transmission despite the pathogen needing to overcome different immunological systems and the fitness costs to vectors.
However, where vectors are present, there is a significant risk of pathogen emergence or adaptation. Therefore, vector-borne diseases pose a high risk, requiring focused surveillance.

The model included a growth rate of pathogens within hosts and vectors in which diminishing transmission rate over virulence levels are crucial. This is biologically plausible since the pathogen will grow in hosts and vectors where there is a finite number of potential cells or tissues to infect. Previous theoretical frameworks have also considered diminishing returns \citep{dieckmann_adaptive_2002}. 
The tradeoff theory has been discussed by several authors including \cite{ewald_evolution_2004}, and a model formulation by 
\cite{bolker_transient_2009}. Recent empirical observations also pointed to diminishing rates for virus titers in SARS-CoV-2 \citep{mautner_replication_2022} in patients and DENV-2 in mosquitoes \citep{salazar_dengue_2007}. 
Nevertheless, several authors have also examined the challenges of reproducing such effects in various disease contexts, both experimentally and observationally.
\cite{alizon_virulence_2009} provide a review discussing why it is often challenging to evaluate the tradeoff hypothesis empirically.
In this aspect,
\cite{doumayrou_experimental_2013} found evidence for the tradeoff hypothesis with plant pathogens, and 
\cite{jensen_empirical_2006} found evidence in experiments with a castrating bacterium.
It will be highly valuable if more experimental studies focus on evaluating such conditions for vector-borne diseases.

The initial formulation of the model is quite general to provide intuition on the effects of parameters on the basic reproduction number. However, a few assumptions were applied to obtain some of the theoretical results. 
Typically, the model assumes that the distributions of general mortality and recovery rates follow an exponential pattern. This assumption simplifies the model, making it easier to analyze and draw meaningful conclusions.
%
%
The formula for the basic reproduction number in this study is arguably more general than the classical MacDonald and Ross formulation \citep{smith_ross_2012}, which is typically used in traditional models for diseases such as malaria and not intended to investigate evolution of virulence. While the derivation of the reproduction number in these classical models is very useful for analyzing effects such as reductions due to control measures, it is less flexible for studying virulence evolution. However, similarities can still be drawn between the present formulation and more recent ones, such as those proposed by \cite{pinho_modelling_2010} for the study of dengue virus transmission.
Furthermore, the model is sufficiently general to capture transmission cycles of various pathogens, beyond arboviruses and {\it Plasmodium}, which use mosquito species as vectors.


Future studies may explore, within the current framework, the impact of parasite clearance due to immunological effects. The rate of parasite clearance could be integrated into the transmission model as a factor that slows down transmission. This immunological component is expected to become relevant sometime after infection, as its effects are minimal during the initial phase. For example, 
\cite{poehler_estimating_2022} demonstrate the decay of antibodies over time in individuals infected with SARS-CoV-2. Similar antibody decay rates have been observed for DENV \citep{vaughn_dengue_2000}. While a simple model can yield some insights, subsequent research should consider a more complex model that incorporates immunological factors.


The transmission rate formulation includes two essential components for evolutionary directions: the parasite growth rate and the threshold value of parasites. Additional avenues for analyzing virulence can be explored, such as investigating the threshold level of parasites required to cause mortality. The approach taken in this study is likely to improve understanding of virulence impacts, as these rates affect the exponential growth of parasites and can significantly influence mortality over time. Furthermore, an increase in the threshold may be attributed to the host lacking immunological adaptation.





Increased surveillance is essential for vector-borne diseases. The GBD project indicates several of these neglected diseases, such as leishmaniasis and yellow fever. Pandemic preparedness demands timely data collection and monitoring, especially for currently overlooked diseases. Variability in climate conditions may disrupt existing systems and lead to increased virulence, even if temporary, which can still result in severe cases.
The theoretical framework in this study is valuable for advancing knowledge,  comparing different scenarios, and assessing risks in changing environments.

%










\section*{Acknowledgments}

DAMV is grateful for the National Council for Scientific and Technological Development (CNPq/Brazil, Ref. 312282/2022-2), Fundação Carlos Chagas Filho de Amparo à Pesquisa do Estado do Rio de Janeiro (Ref. E-26/204.108/2024), and CAPES (Service Code 001). DAMV is grateful to the Center for Health and Wellbeing (CHW) at Princeton University, as most of this work was done during his time as Visiting Research Scholar at CHW.


\bibliography{EvolutionInfectious2,Evol_others}

\newpage





{\bf \Large Appendix}

\subsection*{Complete Model}

The complete model is given by a system of Ordinary Differential Equations (ODE) that describe two coupled SEIRD systems composed of hosts and vectors.  Variables $S_x$, $E_x$, $I_x$, and $R_x$ describe the number of susceptible, exposed, infected, and recovered individuals, where $x={h,v}$, depending on vector/host ($v/h$):
\begin{align*}
    \frac{dS_h}{dt} & = \mu H - m \beta_h(\delta, \omega) S_h I_v/M - \mu S_h\\
    \frac{dE_h}{dt} & = m \beta_h(\delta, \omega) S_h I_v/M - (\theta_h + \mu) E_h \\
    \frac{dI_h}{dt} & = \theta_h(\omega) E_h - (\gamma_h(\omega) + \omega + \mu) I_h\\
    \frac{dD_h}{dt} & =  \omega I_h\\
    \frac{dR_h}{dt} & = \gamma_h(\omega) I_h - \mu R_h\\
    \frac{dS_v}{dt} & = g M -\beta_v(\delta, \omega) S_v I_h/H - g S_v\\
    \frac{dE_v}{dt} & = \beta_v(\delta, \omega) S_v I_h/H - (\theta_v(\delta) + g) E_v \\
    \frac{dI_v}{dt} & = \theta_v(\delta) E_v - (\gamma_v(\delta) + \delta + g) I_v\\
    \frac{dD_v}{dt} & =  \delta I_v\\
    \frac{dR_v}{dt} & = \gamma_v(\delta) I_v - g R_v
\end{align*}
where $m = M/H$.

Normalized equations: 
%
\begin{align*}
    \frac{dS_h}{dt} & = \mu  - m \beta_h(\delta, \omega) S_h I_v - \mu S_h\\
    \frac{dE_h}{dt} & = m \beta_h(\delta, \omega) S_h I_v - (\theta_h + \mu) E_h \\
    \frac{dI_h}{dt} & = \theta_h(\omega) E_h - (\gamma_h(\omega) + \omega + \mu) I_h\\
    \frac{dD_h}{dt} & =  \omega I_h\\
    \frac{dR_h}{dt} & = \gamma_h(\omega) I_h - \mu R_h\\
    \frac{dS_v}{dt} & = g -\beta_v(\delta, \omega) S_v I_h - g S_v\\
    \frac{dE_v}{dt} & = \beta_v(\delta, \omega) S_v I_h - (\theta_v(\delta) + g) E_v \\
    \frac{dI_v}{dt} & = \theta_v(\delta) E_v - (\gamma_v(\delta) + \delta + g) I_v\\
    \frac{dD_v}{dt} & =  \delta I_v\\
    \frac{dR_v}{dt} & = \gamma_v(\delta) I_v - g R_v
\end{align*}

\subsection*{Finding the basic reproduction number}

The matrix generation method is applied to find the basic reproduction number.
The conditions for a disease-free state are that $S_v = M$ and $S_h = H$.
Under these conditions the system of equations related to disease states corresponding to exposed and infected individuals are: 
%
\begin{align*}
    \frac{dE_h}{dt} & = m \beta_h(\delta, \omega) I_v - (\theta_h + \mu) E_h \\
    \frac{dI_h}{dt} & = \theta_h(\omega) E_h - (\gamma_h(\omega) + \omega + \mu) I_h\\
    \frac{dE_v}{dt} & = \beta_v(\delta, \omega) I_h - (\theta_v(\delta) + g) E_v \\
    \frac{dI_v}{dt} & = \theta_v(\delta) E_v - (\gamma_v(\delta) + \delta + g) I_v\\
\end{align*}

This system permits to obtain the matrices $T$ and $\Sigma$ which correspond to the decomposition into transmission and transition parts, respectively, of the equations. 
\begin{equation*}
T = 
\begin{pmatrix}
    0 & 0 & 0 & m \beta_h\\
    0 & 0 & 0 & 0 \\
    0 & \beta_v & 0 & 0 \\
    0 & 0 & 0 & 0
\end{pmatrix} \mbox{, and}
\end{equation*}

\begin{equation*}
\Sigma =     
\begin{pmatrix}
    -(\theta_h+\mu) & 0 & 0 & 0\\
    \theta_h & -(\gamma_h(\omega) + \omega + \mu) & 0 & 0 \\
    0 & 0 & -(\theta_v(\delta) + g) & 0 \\
    0 & 0 & -\theta_v(\delta) & -(\gamma(\delta) + \delta + g)
\end{pmatrix}.
\end{equation*}

The inverse of matrix $\Sigma$ is a step to obtain the critical matrix given by $-T \Sigma^{-1}$ as follows:
\begin{equation*}
\Sigma^{-1} =     
\begin{pmatrix}
    \frac{- 1}{\theta_h+\mu} & 0 & 0 & 0\\
    \frac{\theta_h}{(\theta_h+\mu) (\gamma_h(\omega) + \omega + \mu)}& \frac{- 1}{\gamma_h(\omega) + \omega + \mu} & 0 & 0 \\
    0 & 0 & \frac{-1}{\theta_v(\delta) + g} & 0 \\
    0 & 0 & \frac{\theta_v(\delta)}{(\theta_v(\delta) + g) (\gamma(\delta) + \delta + g)} & \frac{- 1}{\gamma(\delta) + \delta + g}
\end{pmatrix}
\end{equation*}
\begin{equation*}
-T \Sigma^{-1} =     
\begin{pmatrix}
    0 & 0 & \frac{\theta_v(\delta) m \beta_h }{(\theta_v(\delta) + g) (\gamma(\delta) + \delta + g)} & \frac{- m \beta_h}{\gamma(\delta) + \delta + g} \\
    0 & 0 & 0 & 0\\ \frac{\theta_h \beta_v}{(\theta_h+\mu) (\gamma_h(\omega) + \omega + \mu)}& - \frac{\beta_v }{\gamma_h(\omega) + \omega + \mu} & 0 & 0\\
    0 & 0 & 0 & 0
\end{pmatrix}
\end{equation*}

Matrix $-T \Sigma^{-1}$ is critical because its spectral radius gives the basic reproduction number: 
\begin{equation}
R_0 = \sqrt{\frac{\theta_v(\delta) m \beta_h }{((\theta_v(\delta) + g) (\gamma(\delta) + \delta + g))} \frac{\theta_h \beta_v}{((\theta_h+\mu) (\gamma_h(\omega) + \omega + \mu))} }.
\label{eq:R0appendix}
\end{equation}

\subsection*{Theoretical optimal conditions for maximising the basic reproduction number}





The basic reproduction number is maximized when the partial derivatives 
with respect to vector mortality and host mortality equal zero. Hence, this necessary and sufficient condition permits to find the optimal mortality rates as follows (Obs:maximizing $R_0^2$).
%
%
\begin{align}
\label{eq:optstates1}
\frac{\partial_\omega \theta_\omega(\omega)}{\theta_\omega(\omega)} + \frac{\partial_\omega \beta_h(\delta, \omega)}{\beta_h(\delta, \omega)} + \frac{\partial_\omega \beta_m(\delta, \omega)}{\beta_m(\delta, \omega)} & = \frac{\partial_\omega \theta_h(\omega)}{\mu + \theta_h(\omega)} + \frac{\partial_\omega \gamma_h(\omega) + 1}{\mu + \gamma_h(\omega) + \omega}    \\ \label{eq:optstates2}
\frac{\partial_\delta \theta_\delta(\delta)}{\theta_\delta(\delta)} +\frac{\partial_\delta \beta_h(\delta, \omega)}{\beta_h(\delta, \omega)} + \frac{\partial_\delta\beta_m(\delta, \omega)}{\beta_m(\delta, \omega)} & = \frac{\partial_\delta \theta_v(\omega)}{\mu + \theta_v(\delta)} + \frac{\partial_\delta \gamma_v(\delta) + 1}{\mu + \gamma_v(\delta) + \delta}    
\end{align}

The left--hand side of Equations \ref{eq:optstates1} and \ref{eq:optstates2} contains the logarithmic derivative 
of functions $\beta_h(\omega, \delta)$ and $\beta_v(\omega, \delta)$. It is useful to apply the operator $L$, defined by $Lu(x) = \frac{\frac{du}{dx}}{u(x)}$ for function $u(x)$ , and a fraction of $A$ =$f_A$.
The optimal level of virulence $\omega$ is given by the solution of Equation: 
%
$$
L_\omega \theta_\omega(\omega) +
 L_{\omega} \beta_h(\delta, \omega) + L_{\omega} \beta_m(\delta, \omega) = \frac{\partial_\omega \theta(\omega)}{\mu + \theta(\omega)} + \frac{\partial_\omega \gamma_h(\omega) + 1}{\mu + \gamma_h(\omega) + \omega}.
$$

\pagebreak

\subsection*{Evaluation of Case Fatality Ratios for countries' populations}

The Global Burden of Disease (GBD) project regularly collects data on various health conditions, including infectious diseases \citep{institute_of_health_metrics_and_evaluation_global_2020}. The most recent data, as of 2019, provides information on the number of cases and deaths for diseases listed in the Supplementary Text.
The study included 169 cause names for diseases and health conditions that might lead to death.  A number of 30 infectious diseases were selected from this list with a simple criterion that should involve a pathogen for transmission.   Diseases Cisticercosis, nematode infections, Schistosomiasis, and "other neglected diseases" did not include values for either incidence or mortality and were excluded.
The World Bank's classification of countries per income level was used to categorize countries reported in the study. The final list had 22 infectious diseases or groups of infectious diseases.
The case fatality ratio (CFR) of a disease in a given year is calculated in this study by taking the ratio between the incidence and mortality 
of the disease in that year \citep{lash_modern_2021}. This ratio has the advantage of not requiring population estimates for countries.

The three most recent GBD studies were conducted in the years 2019, 2010, and 2000. The fatality rates were observed from empirical data to identify diseases with fatality rates above 1\% in recent years. 
The goal of the original GBD study, however, was not to compare fatality rates among diseases, and some of these diseases may have effective treatment options, such as vaccines.
%


\begin{figure}[h!]
\centering
\includegraphics[width=0.90\linewidth]{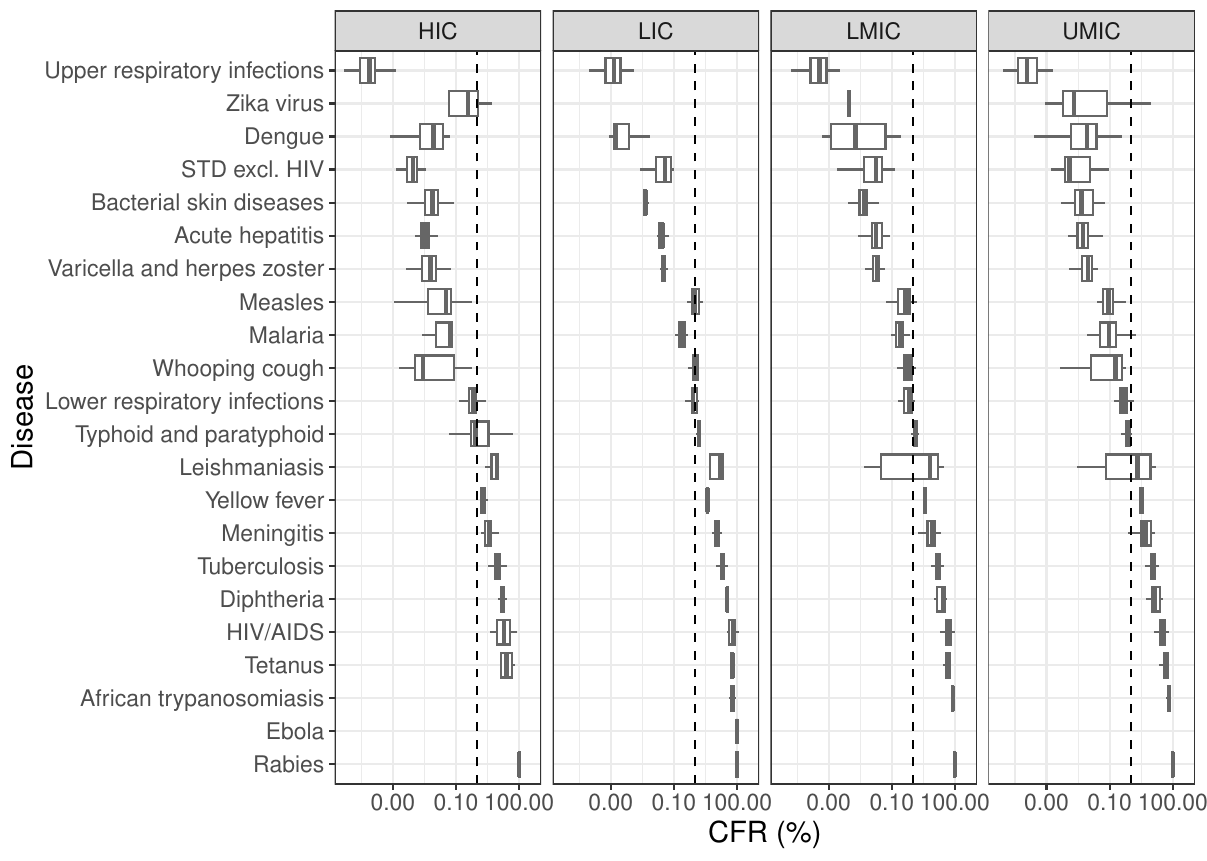}
\caption{Case fatality rate distribution accross countries evaluated in the GBD study. CFR values are in logarithmic scale for better visualization purposes. HIC = High Income Country, LIC = Low Income Country, LMIC = Lower Middle Income Country, UMIC = Upper Middle Income Country.  STD excl. HIV = Sexually transmitted diseases excluding HIV.  Information low--income countries had information about ebola. African trypanosomiasis information was only available for LIC, LMIC and UMIC countries.}
\label{fig:gbd}
\end{figure}

Figure \ref{fig:gbd} shows that
the diseases from the GBD study \citep{institute_of_health_metrics_and_evaluation_global_2020} with higher CFR include vector-host diseases or might have multiple hosts such as schistosomiasis and rabies. 
As expected, the income level of countries has a noticeable impact, with a higher proportion of diseases exceeding a 1\% fatality rate in low-income countries compared to high-income countries.
Still, several observations apply to all categories.
The variability of case fatality rates of diseases from dengue to leishmaniasis, as listed in Figure \ref{fig:gbd} is very high.
Median values for 
leishmaniasis, Yellow fever, meningitis, diphteria, tetanus, trypanosomiasis, rabies and schistosomiasis are above 1\% for High income countries.
The intervals exhibited for Yellow fever, meningitis, diphteria, tetanus, trypanosomiasis, rabies, and schistosomiasis show that above 95\% of records are above 1\%. 
Trypanosomiasis, rabies and schostosomiasis have extremely high fatality rates. 

The number of cases and deaths from the Global Burden of Disease (GBD) study from 2019 permitted an evaluation of case fatality rates, which indeed show high fatality rates for diseases such as trypanosomiasis, yellow fever, and leishmaniasis. However, comparing fatality rates for different diseases within this dataset has problems due to biases. Some of these diseases have well-established treatment schemes, and some might even have vaccines available. Additionally, countries have highly heterogeneous healthcare systems that might also impact the outcomes for infected individuals.
%
The main purpose of analyzing fatality rates using the GBD data is to highlight the variability in case fatality rates and to distinguish those with high fatality rates.
The original goal of the GBD study is to evaluate the burden of disease, and drawing comparisons on disease fatality rates requires caution.
%
First, there is likely a strong bias in reporting for some diseases, especially when only deaths are reported, which overestimates the CFR. Second, there are very heterogeneous levels of surveillance and treatment across countries, from low-income to high-income countries. More resources can be used in the surveillance, prevention, and treatment of infectious diseases if they are available, which highly depends on the countries. The analysis in categories recognizes such effects, although the current categorization still groups multiple, possibly heterogeneous, countries into the same group.
Third, the diseases listed in the analyses included diseases with first-line drug treatments and diseases for which vaccines are regularly produced, whereas others, such as rabies, currently do not have treatment. Again, the availability of treatments and vaccines might depend on the countries' income levels. Also, some diseases might exhibit high variability due to low numbers. Yellow fever in Brazil is an example, with years of very high CFR (35\%) \citep{douam_yellow_2018}, followed by years with few cases and highly variable CFR.
Nevertheless, the analysis conducted here provides evidence that fatality rates across such comprehensive set of diseases vary significantly, as expected. Importantly, the inclusion of a disease in the GBD study indicates that it has a substantial burden, which means that the final list of diseases presented here has a non-negligible case fatality rate (CFR).


\subsection*{List of evaluated diseases in GBD study}

All
diseases and conditions in GBD study:  Urinary diseases and male infertility, Exposure to forces of nature, Environmental heat and cold exposure, Ebola, Executions and police conflict, Eating disorders, Diabetes mellitus, Acute glomerulonephritis, Chronic kidney disease, Gynecological diseases, Bacterial skin diseases, Upper digestive system diseases, Esophageal cancer, Stomach cancer, Paralytic ileus and intestinal obstruction, Inguinal, femoral, and abdominal hernia, Inflammatory bowel disease, Vascular intestinal disorders, Gallbladder and biliary diseases, Pancreatitis, Other digestive diseases, Falls, Drowning, Fire, heat, and hot substances, Poisonings, Exposure to mechanical forces, Other unspecified infectious diseases, Liver cancer, Larynx cancer, Tracheal, bronchus, and lung cancer, Breast cancer, Cervical cancer, Meningitis, Encephalitis, Diphtheria, Whooping cough, Tetanus, Measles, Varicella and herpes zoster, Malaria, Chagas disease, Leishmaniasis, African trypanosomiasis, Schistosomiasis, Cysticercosis, Cystic echinococcosis, Decubitus ulcer, Other skin and subcutaneous diseases, Sudden infant death syndrome, Road injuries, Other transport injuries, Tuberculosis, HIV/AIDS, Diarrheal diseases, Other intestinal infectious diseases, Lower respiratory infections, Upper respiratory infections, Otitis media, Testicular cancer, Kidney cancer, Bladder cancer, Brain and central nervous system cancer, Endocrine, metabolic, blood, and immune disorders, Rheumatoid arthritis, Other musculoskeletal disorders, Congenital birth defects, Protein-energy malnutrition, Other nutritional deficiencies, Sexually transmitted infections excluding HIV, Acute hepatitis, Appendicitis, Other neglected tropical diseases, Maternal disorders, Neonatal disorders, Endocarditis, Non-rheumatic valvular heart disease, Chronic obstructive pulmonary disease, Pneumoconiosis, Prostate cancer, Colon and rectum cancer, Lip and oral cavity cancer, Nasopharynx cancer, Other pharynx cancer, Gallbladder and biliary tract cancer, Pancreatic cancer, Malignant skin melanoma, Non-melanoma skin cancer, Zika virus, Conflict and terrorism, Uterine cancer, Multiple sclerosis, Motor neuron disease, Other neurological disorders, Alcohol use disorders, Drug use disorders, Alzheimer's disease and other dementias, Parkinson's disease, Idiopathic epilepsy, Adverse effects of medical treatment, Animal contact, Foreign body, Other unintentional injuries, Asthma, Interstitial lung disease and pulmonary sarcoidosis, Other chronic respiratory diseases, Cirrhosis and other chronic liver diseases, Typhoid and paratyphoid, Invasive Non-typhoidal Salmonella (iNTS), Hypertensive heart disease, Cardiomyopathy and myocarditis, Atrial fibrillation and flutter, Aortic aneurysm, Peripheral artery disease, Ovarian cancer, Thyroid cancer, Mesothelioma, Self-harm, Non-Hodgkin lymphoma, Multiple myeloma, Leukemia, Other neoplasms, Rheumatic heart disease, Interpersonal violence, Ischemic heart disease, Stroke, Other malignant neoplasms, Other cardiovascular and circulatory diseases, Hemoglobinopathies and hemolytic anemias, Hodgkin lymphoma, Dengue, Yellow fever, Rabies, Intestinal nematode infections, Osteoarthritis, Low back pain, Neck pain, Gout, Blindness and vision loss, Food-borne trematodiases, Age-related and other hearing loss, Other sense organ diseases, Oral disorders, Autism spectrum disorders, Attention-deficit/hyperactivity disorder, Scabies, Fungal skin diseases, Viral skin diseases, Acne vulgaris, Alopecia areata, Pruritus, Urticaria, Conduct disorder, Idiopathic developmental intellectual disability, Other mental disorders, Lymphatic filariasis, Onchocerciasis, Trachoma, Iodine deficiency, Vitamin A deficiency, Dermatitis, Psoriasis, Bipolar disorder, Anxiety disorders, Dietary iron deficiency, Headache disorders, Leprosy, Schizophrenia, Depressive disorders, Guinea worm disease \\

Diseases with pathogen: Ebola, Bacterial skin diseases, Meningitis, Diphtheria, Whooping cough, Tetanus, Measles, Varicella and herpes zoster, Malaria, Leishmaniasis, African trypanosomiasis, Schistosomiasis, Cysticercosis, Tuberculosis, HIV/AIDS, Lower respiratory infections, Upper respiratory infections, Sexually transmitted infections excluding HIV, Acute hepatitis, Other neglected tropical diseases, Zika virus, Typhoid and paratyphoid, Dengue, Yellow fever, Rabies, Intestinal nematode infections, Scabies, Lymphatic filariasis, Trachoma, Guinea worm disease 

  Final set of diseases in evaluation: Bacterial skin diseases, Meningitis, Diphtheria, Whooping cough, Tetanus, Measles, Varicella and herpes zoster, Tuberculosis, HIV/AIDS, Lower respiratory infections, Upper respiratory infections, Sexually transmitted infections excluding HIV, Acute hepatitis, Leishmaniasis, Typhoid and paratyphoid, Rabies, Malaria, Dengue, Yellow fever, African trypanosomiasis, Zika virus, Ebola

\end{document}